\documentstyle[12pt,epsfig]{article}
\newfont{\Bbb}{msbm10 scaled 1200}     
\newcommand{\mathbb}[1]{\mbox{\Bbb #1}}

\def\TL{\hfil$\displaystyle{##}$}
\def\TR{$\displaystyle{{}##}$\hfil}

\def\lbldef#1#2{\expandafter\gdef\csname #1\endcsname {#2}}
\def\eqn#1#2{\lbldef{#1}{(\ref{#1})}%
\begin{equation} #2 \label{#1} \end{equation}}
\def\eqalign#1{\vcenter{\openup1\jot
    \halign{\strut\span\TL & \span\TR\cr #1 \cr
   }}}

\def\href#1#2{#2}

\newcommand{\beq}{\begin{equation}}
\newcommand{\eeq}{\end{equation}}

\newcommand{\ber}{\begin{eqnarray}}
\newcommand{\eer}{\end{eqnarray}}

\newcommand{\beqar}{\begin{eqnarray}}

\newcommand{\eeqar}{\end{eqnarray}}

\newcommand{\ba}{\begin{eqnarray}}
\newcommand{\ea}{\end{eqnarray}}

\newcommand{\dsl}
  {\kern.06em\hbox{\raise.15ex\hbox{$/$}\kern-.56em\hbox{$\partial$}}}

\newcommand{\eeqarr}{\end{eqnarray}}
\newcommand{\ZZ}{{\rm \kern 0.275em Z \kern -0.92em Z}\;}

\begin{document}
\baselineskip=15.5pt
\pagestyle{plain}
\setcounter{page}{1}
\begin{titlepage}

\leftline{\tt hep-th/0008001}

\vskip -.8cm


\begin{center}

\vskip 1.7 cm

{\LARGE Towards  the large $N$ limit }
\vskip .3cm 
{\LARGE  of pure ${\cal N}=1$ 
Super Yang Mills }

\vskip 1.5cm
{\large 
Juan Maldacena  and Carlos Nu\~nez
}
\vskip 1.2cm

Jefferson  Laboratory of Physics,
Harvard University, Cambridge, MA  02138, USA

\vskip 0.5cm

\vspace{1cm}

{\bf Abstract}

\end{center}
We find the gravity solution corresponding to a large number of 
 NS or D fivebranes
wrapped on a two sphere so  that we have pure 
${\cal N}=1$ super Yang-Mills in the IR. The supergravity 
solution is smooth, it shows confinement
 and it breaks the $U(1)_R$ chiral symmetry in the appropriate
way.  When the gravity approximation is valid the masses of glueballs
are comparable to the masses of Kaluza Klein states on the fivebrane,
but if we could quantize strings on this background it looks like
we should be able 
to decouple the KK states.

\noindent

\end{titlepage}

\newpage


\section{Introduction}
\label{intro}

The AdS/CFT correspondence
\cite{Maldacena:1998re,Gubser:1998bc,Witten:1998qj}
 gives the large $N$ dual description for
${\cal N} =4$ super Yang Mills. It would be nice to 
find similar correspondences
for pure Yang-Mills theories with less supersymmetry. 
By ``pure'' we mean without
matter. In this paper we make some progress in this direction by finding 
a geometry that is dual to a little string  theory that reduces to pure ${\cal N} =1 $ 
super Yang-Mills in the IR. 
We consider a little string theory \cite{Berkooz:1997cq},
 or NS 5 brane theory, in type IIB string 
theory. In the IR this theory reduces to six dimensional super Yang Mills with 
sixteen supercharges. We wrap this brane on $S^2$ and we twist the normal bundle 
in such a way that we preserve only 1/4 of the supersymmetries and we give a mass
to the four  scalar fields. This theory reduces then to pure ${\cal N} =1$
super Yang-Mills in the IR. 
Starting from the geometric description of the NS 5 brane theory, we modify the 
boundary conditions for the metric to take into account that we are wrapping 
an $S^2$ and we have an appropriately twisted normal bundle. Then we find the 
supergravity solution using methods similar to the one in 
\cite{Maldacena:2000mw},
 i.e. 
reducing the problem to gauged supergravity in seven dimensions, 
reading the 
solution from \cite{Chamseddine:1997nm} (see also
\cite{Chamseddine:1998mc}, for further studies on the solution) 
and then lifting it up to ten dimensions using 
\cite{Cvetic:2000dm, Chamseddine:1999uy}.
When the supergravity approximation is valid the little string theory  scale and the 
scale of the four dimensional theory are comparable. Nevertheless the solution
has all the expected qualitative features. 
It has a $U(1)_R$ symmetry broken in the UV to 
$Z_{2N}$ and the full solution breaks it further to $Z_2$ and 
we find $N$ different
solutions. The theory is confining and it is magnetically screening. 
It has domain walls between the different vacua.
Strings can end on the domain walls. 

When we try to take the decoupling limit we find a rather 
precise RR $\sigma$ model that
we should quantize in order to find the decoupled string 
theory describing ${\cal N} =1$
super Yang-Mills, though some aspects are a bit unclear at this stage. 
It is not clear, for example, how the string coupling is quantized. 

While this paper was in preparation we became aware of 
 the work of I. Klebanov and M. Strassler
\cite{Klebanov:2000hb}
 which has some overlap with ours since 
they also study a theory that reduces to
pure ${\cal N} =1$ Yang-Mills in the IR. 
We also became aware of a paper by C. Vafa on the same
topic  that appeared after the first version of this work
\cite{Vafa:2000wi}.
We thank them  for discussions and for 
 telling us about the relevance of the deformed
conifold metric.

\section{NS 5-branes on $S^2$}

Since the appropriate UV description of the fivebrane theory is
the little string theory, or NS-5brane, we start with an NS 5-brane 
in type IIB string theory. 
The geometry dual to the little string theory is 
\eqn{flat}{\eqalign{
ds^2_{str}  =& dx_6^2 + N ( d \rho^2 + d\Omega^2_3 )
\cr
e^\phi = & e^{\phi_0 - \rho}
}}
where $\phi_0$ is an arbitrary constant that can be changed by shifting
$\rho$. $N$ is the number of fivebranes. 
 This represents a fivebrane whose worldvolume is $R^6$. 
Now we would like to consider a brane whose worldvolume is 
\eqn{wvol}{
ds_6^2 = dx_4^2 + N e^{2 g} d\Omega^2_2
}
 so that the brane is wrapped
on a two sphere of radius $R^2 = N e^{2 g}$. 
The factor of $N$ is introduced just for later convenience. 
In order to preserve supersymmetry we should twist the normal bundle. 
This twisting is achieved by embedding the spin connection into  the 
R-symmetry group.
Since the non-trivial part of the spin connection is in a $U(1)$ subgroup
we should choose how to embedd the $U(1)$ in 
$SO(4) \sim SU(2)_R \times SU(2)_L$. $SO(4)$ is the R-symmetry group 
of the NS fivebrane and it rotates the $S^3$ in \flat .
If we embedd the spin connection in $U(1)_R \subset SU(2)_R \subset SO(4)$
we preserve only four supercharges or ${\cal N}=1$ supersymmetry in four
 dimensions.
Let us see this more precisely from the  fivebrane worldvolume 
point of view. 
The spinors that generate the supersymmetries on the NS fivebrane 
are two six dimensional spinors with positive chirality that are in the ${\bf (2,0)}$
of $SU(2)_R\times SU(2)_L$ and two negative chirality spinors in the 
${\bf (0,2)}$. 
 The supersymmetries that are generated
by the spinors transforming under $SU(2)_L$ are broken.
The preserved supersymmetries have positive chirality in six dimensions
and are such that  the $U(1)_R $ charge is correlated
with the chirality of the spinor in the two directions of the 
sphere. We see that this leaves us with 1/4 of the original 
supersymmetries of the five-brane. 
The four scalars transverse to the fivebrane transform under the 
${ \bf (2,2)}$ of $SU(2)_R \times SU(2)_L$. This implies that, after twisting,  
they become   spinors on the two sphere so 
that they do not  have any zero modes. In the IR the only massless
fields are the gauge fields and the gauginos. So in the IR we have 
pure ${\cal N} = 1$ super Yang-Mills. The value of the Yang-Mills 
coupling is given in terms of the volume of the sphere by
\eqn{coup}{
 { 1 \over g_4^2 } = {Vol_{S^2}  \over g_6^2 } = { N e^{2 g} \over 2 \pi^2 } ~ .
}
We have described this in terms of the low energy field theory on 
the fivebrane so we are implicitly assuming that the volume of
the $S^2$ is much larger than the five dimensional gauge coupling so that
we can apply the low energy description for the fields on the fivebrane.
We will later discuss more precisely  the limit in which the 
four dimensional super Yang-Mills theory is expected to decouple. 
For the moment we will analyze this fivebrane theory, without taking the
decoupling limit. 
This twisted fivebrane theory seems to have a $U(1)_R$ symmetry which is the 
$U(1)_R$ that we are twisting. This  is the $U(1)_R$ symmetry of
${\cal N} =1$ super Yang-Mills, it acts on the gluinos but not on the gauge fields.
We will see that this $U(1)_R$ symmetry is broken to $Z_{2N}$ by
worldsheet instantons
in the NS description.  
This twisting also preserves the $SU(2)_L$ symmetry of the fivebrane theory. But this
symmetry does not act on the massless fields, it only acts on the Kaluza Klein modes
which are expected to decouple in the IR. 

\section{Finding the gravity solution}

As explained in \cite{Maldacena:2000mw} we need to impose an appropriate
boundary 
condition for the geometry. In this case the
boundary is at $\rho \to \infty$. So we impose the condition that the 
seven dimensional geometry has a boundary which is $R^4 \times  S^2$ and
we implement the twisting by imposing appropriate boundary conditions
for the seven dimensional gauge fields which come from the isometries of
$S^3$. In this case we will impose that the $U(1)_R$ gauge field in 
$SU(2)_R$ is equal to the spin connection on the $S^2$.In other
words, we set $A^3 = \cos\theta d \varphi $ for large $\rho$. It
turns out that an ansatz like this is possible only if we allow the
volume of $S^2$ to grow as $\rho \to \infty$. This is related to
the 
running of the coupling in four dimensions. 
The string frame geometry will involve only the six dimensions 
parametrized by $\rho$, the two-sphere and the coordinates on the three sphere. 
In fact this solution is a particular case of the general solutions analyzed in 
\cite{Strominger:1986uh}. 
It is a ``compactification'' with torsion to four dimensions. It is not
really a compactification because the four dimensional Newton's constant is zero 
in our case since we are decoupling gravity\footnote{In \cite{Strominger:1986uh}
the heterotic case was
considered, but the results  extend simply to
the type II case.}.
Though the general 
conditions for a supersymmetric compactification were 
completely spelled out in 
\cite{Strominger:1986uh} it is convenient, in order to find an explicit
solution, to use a 
 different technique. 

Since the boundary conditions are imposed on seven dimensional fields it is
convenient to work with seven dimensional gauged supergravity 
\cite{Townsend:1983kk}, which
is a truncation to seven dimensions of the ten dimensional equations in 
the near horizon region  of a five brane
\cite{Cvetic:2000dm,Chamseddine:1999uy}. 
This seven dimensional theory contains the metric, 
a dilaton, $SU(2)_R$ gauge fields
and a  $B_{\mu\nu} $ field, which accounts for the seven dimensional 
components of the $B$ field. In \cite{Townsend:1983kk} the seven dimensional 
$B_{\mu\nu}$ was
dualized into a three form with
a four form field strength. We set this field to zero for the time being.
{}From the discussion above we expect that in string frame this seven dimensional
solution will be $R^4$ times a three dimensional geometry parametrized by 
$\rho$ and the coordinates on $S^2$, with  non-zero components of the 
gauge fields 
only along  these three dimensions. 
We can see this explicitly from the form of the supersymmetry variations in 
seven dimensional string frame 
\eqn{susy}{ \eqalign{
\delta \lambda & = \not D \phi \epsilon - { i\sqrt{N} \over 4 } \Gamma^{\mu \nu} F_{\mu \nu} \epsilon 
+{ 1 \over \sqrt{N}} \epsilon
\cr
\delta \chi_\mu & = (D_\mu + i A_\mu ) \epsilon -
 { i \sqrt{N} \over 2 } F_{\mu \nu} \Gamma^\nu \epsilon
}}
The gauge fields are 
\eqn{gaugedef}{
A_\mu ={ 1\over 2}  \sigma^a A^a_\mu ~,~~~~~~~ F={1 \over 2} F^a \sigma^a ~,~~~~~~~~ F^a_{\mu\nu} = 
\partial_\mu A^a_\nu - \partial_\nu A^a_\mu + \epsilon^{abc} A^b_\mu A^c_\nu 
}
where $\sigma^a$ are the  Pauli matrices and the spinors carry an extra
$SU(2)_R$ index on which these Pauli matrices act. 
Let us first try a simple ansatz which correctely describes the UV (or large $\rho$) 
part of the solution
\eqn{simpleans}{\eqalign{
ds^2_{str} = & dx^2_4 + N[ d\rho^2 +  e^{2 g(\rho) }(d \theta^2 + \sin^2 \theta d\varphi^2 )] 
\cr 
A^3 = & \cos \theta d\varphi
}}
and all other gauge fields equal to zero. 
We see that this ansatz respects the boundary conditions that we want to impose at
infinity.
Putting this in the supersymmetry equations \susy\ we find the  solution
\eqn{simplesol}{
e^{2 g(\rho)} = \rho ~,~~~~~~~~ \phi = \phi_0 - \rho + { 1 \over 4} \log \rho 
}
We have absorbed the integration constant in the definition of $\phi_0$. 
The dependence of $g$ on $\rho$ is related to the dependence of the four
dimensional coupling \coup\ on the scale $ \rho$.

This metric \simpleans\ is singular at $\rho=0$ and the singularity 
is of a bad type according to
the criteria in \cite{Gubser:2000nd, Maldacena:2000mw}.
The necessary ingredient in order to resolve the singularity comes in when we consider
the symmetries of this solutions. The metric  we found still has the $U(1)_R$ symmetry,
these are $U(1)$ charge rotations in the $\sigma^3$ directions
 in this seven dimensional description.
We expect, however
that this symmetry should be broken by the choice of vacuum in the four dimensional 
gauge theory. 
Naively we expect that the solution should be such that the $S^2$ should shrink to 
zero. A similar effect (actually, the opposite)  was found in the 
topological string theory/Chern Simons correspondence in 
\cite{Gopakumar:1998ki} where one
starts with a large $N$ 
Chern Simons theory on $S^3$, where the $S^3$ is that of a resolved conifold, 
and one ends 
with a dual geometry which is  a conifold resolved to a finite size $S^2$ \footnote{
This point was emphasized to us  by C. Vafa.}.
In our case we cannot shrink the $S^2$ to zero because there is a non-trivial $U(1) $ flux
through it. If we view this bundle as an $SU(2)$ bundle then it becomes clear that
we can first trivialize the bundle and then shrink the $S^2$ to zero. 
Actually, this problem is completely analogous to a magnetic monopole in $SU(2)$ theory
vs. the Dirac monopole. The solutions we found here is analogous to a Dirac monopole 
 \cite{Chamseddine:1997nm}\cite{Chamseddine:1998mc}. 
So we should look for the solution analogous to the $SU(2)$ monopole, which will have
$A^1, A^2$ non-vanishing. These fields are charged under $U(1)_R$ and will 
  thus break the $U(1)$ symmetry.   
Fortunately this solution was found in \cite{Chamseddine:1997nm}.
Actually, the solution  in \cite{Chamseddine:1997nm} is 
 a monopole-like solution of a four dimensional gauged supergravity.  In order to see that 
it is the same as the solution we want we should write the supersymmetry variation 
 equations of 
\cite{Chamseddine:1997nm}
 in string frame. Their solution only involves three of the dimensions
and the susy equations are the same as ours in those three dimensions. 
So we can simply read off their solution \cite{Chamseddine:1997nm} 
\eqn{solseven}{\eqalign{
A =& { 1 \over 2} \left[ \sigma^1 a(\rho) d \theta + \sigma^2 a(\rho) \sin\theta d\varphi +
\sigma^3 \cos\theta d \varphi \right]
\cr
a(\rho) = &{ 2 \rho \over \sinh 2 \rho}
\cr
e^{2 g}  =& \rho \coth 2\rho - { \rho^2 \over \sinh^2 2 \rho } - { 1 \over 4}
\cr
e^{ 2 \phi}  =& e^{2 \phi_0} {2 e^{g} \over \sinh 2 \rho}
}}
We see that for large $\rho$ these functions go as 
$ e^{2 g } \sim \rho $, $ a \sim o(e^{-2 \rho}) $ and the dilaton also has the 
same behaviour as in the previous $U(1)$ solution. This implies that the solution
has the proper UV behaviour. 
At the origin $ \rho =0$ the metric goes as $e^{2 g} \sim \rho^2$ so that the metric
is non-singular. It is also easy to check that $A$ is pure gauge at the origin.
In fact this can be seen through the following  gauge transformation 
\eqn{gauge}{
h = e^{i \sigma^1 \theta \over 2} e^{ i \sigma^3 \varphi \over 2}~,~~~~~~ i A = dh h^{-1}
+ o(\rho^2) ~.}
Note that in this monopole-like  solution there is no Higgs field. 

Now that we have found the seven dimensional solution it is possible to lift it 
up to ten dimensions using the formulas in
 \cite{Cvetic:2000dm,Chamseddine:1999uy}.
In order to write the solution it is useful to choose Euler angles on the sphere $S^3$, and
define the left invariant one forms  by viewing
the sphere as the SU(2) group 
\eqn{sph}{\eqalign{
g &= e^{ i \psi \sigma^3 \over 2 } e^{ i \tilde \theta \sigma^1 \over 2 } 
e^{ i \phi \sigma^3 \over 2} 
\cr
 { i \over 2} w^a \sigma^a &  = dg g^{-1} 
\cr
w^1 + i w^2 & = e^{ - i \psi } ( d \tilde \theta + i \sin \tilde \theta d \phi)  ~,~~~~~~~~~~
w^3 = d \psi + cos\tilde \theta  d\phi 
}}

Using the uplifting formulas in \cite{Cvetic:2000dm,Chamseddine:1999uy} 
the ten dimensional
solution is 
\eqn{soltenns}{\eqalign{
ds^2_{str} =& dx_4^2 + N \left[
d \rho^2 + e^{ 2 g(\rho)} ( d \theta^2 + \sin^2  \theta d\varphi^2 ) + {1 \over 4} \sum_a
( w^a - A^a)^2 \right]
\cr
e^{2 \phi} =& e^{ 2 \phi_0}{ 2 e^{g(\rho)} \over \sinh 2 \rho }
\cr
H^{NS} =& N \left[ - {1\over 4} (w^1 -A^1)\wedge (w^2 - A^2) \wedge ( w^3-A^3)  + { 1 \over 4} 
\sum_a F^a \wedge (w^a -A^a) \right] 
}}

The only integration constant in the solution is $\phi_0$, which is the value of the
dilaton at $\rho =0$. 
We see that geometrically the resolution of the singularity 
is the same as  that in \cite{Klebanov:2000hb}. If we wrap branes on the
$S^2$ of a resolved
conifold the twisted field theory on the brane is precisely what we had above and the 
resolution is that the $S^2$ shrinks and the $S^3$ stays with finite size. In fact our
solution is similar to the solution considered in \cite{Klebanov:2000hb} except
that we have
only fractional branes and no regular branes.

\section{ The fate of the $U(1)$ R-symmetry and the $N$ vacua}

Let us  understand why the $U(1)$ symmetry of the solution at infinity is
broken to $Z_{2N}$. In the coordinates we have chosen this $U(1)$ symmetry corresponds
to shifting $ \psi \to \psi + \epsilon$, with $\psi = \psi + 4 \pi $. 
This symmetry is broken by worldsheet instantons. Naively we would say that
instantons of the field theory, which are the strings of the little string theory, are
string worldsheets wrapping $S^2$. This is almost right except that the instanton also 
wraps an $S^2$ inside $S^3$. More precisely, if we parametrize it by 
the coordinates $\theta, ~ \varphi$ of \soltenns\ we also  we have to set
$\tilde \theta = \theta ,~ \phi = \varphi, ~\psi = const$. It is possible to have
a worldsheet with constant $\psi$ thanks to the gauge field $A^3$ since
what appears in the metric is $ d \psi + \cos \tilde \theta d \phi - \cos \theta d \varphi $.
In other words, the coordinate $\psi$ is trivially fibered over the worldsheet so 
 so that
we can pick a configuration with constant $\psi$. 
There will be a flux of the $B$ field over this sphere. This flux however depends on
the point $\psi$ where the sphere is sitting. In fact we can see that, for large $\rho$,  
\eqn{fluxch}{
{1 \over 2 \pi }\int_{\psi_2}  B - {1 \over 2 \pi }\int_{\psi_1}  B = { 1 \over 
2 \pi } \int H d\psi d\theta d \varphi = - N (\psi_2 - \psi_1)
}
So this flux goes as 
\eqn{flux}{
{1 \over 2 \pi }\int_{\psi}  B = b - N \psi
}
This flux is  
 the phase that appears in the worldsheet instanton calculation. This
should be identified with the phase that appears in the field theory 
instanton calculation, which is the field theory $\theta_{FT}$ angle. 
We see here that, as we perform a shift in $\psi$, the phase changes.
This implies that the $U(1)$ symmetry is anomalous, it is changing the field theory
since it is changing the $\theta_{FT}$ angle. By convention we can define the 
field theory theta angle to be the flux at $\psi =0$ and then agree not to change
it by perfoming $U(1)$ transformations. We see, however, 
 that $\theta_{FT}$  is
not changed if we do rotations by $ \psi \to \psi + {2 \pi  n \over N}$, with
$ 0 \leq n < 2 N$. This is precisely the surviving $Z_{2N}$ symmetry in the UV. 
This symmetry is broken to $Z_2$ by the solution  \soltenns\ . The surviving 
$Z_2$ is just $ \psi \to \psi + 2 \pi $ which does not change the solution \soltenns\ .

We should now explain why we have precisely $N$ solutions, or $N$ vacua, for 
each value of $\theta_{FT}$. First we notice that the worldsheet that we were 
talking about around \fluxch\  is contractible in the full geometry. In order to see this
we can bring the sphere close to $\rho =0$ in the geometry \soltenns\ and then 
perform the gauge transformation \gauge , which amounts to a coordinate
transformation
on the three
 sphere. After  this, the worldsheet  is wrapped on the
two sphere that collapses to zero. If the geometry is to be smooth
 the flux on the 
collapsing spherical worldsheet better be a multiple of $2 \pi$. This will not happen
in general. For example, for the solution \soltenns\ we see that if we pick 
a worldsheet wrapping the sphere at $ \psi =0$ and we transport it to the origin, we
do not
 pick any extra flux since the radial components of $H$ in \soltenns\ projected to the
worldsheet worldvolume are  proportional to $\sin \psi$ and we are at $\psi =0$. 
So the flux at the origin is the same as the flux at infinity which in turn is equal 
to $\theta_{FT}$.
So the solution \soltenns\ is a good solution only for  
$\theta_{FT} =0$. Which are the other solutions?. It is easy to see how to generate
new solutions. All we have to do is to rotate the gauge fields by a $U(1)$ transformation
and replace the gauge fields in \soltenns\ by
\eqn{newgauge}{
A' = e^{ i \psi_0 \sigma^3 \over 2} A e^{ -i \psi_0 \sigma^2 \over 2} 
}
This does not change the gauge fields at infinity, so it does not modify the solution
in the UV. 
Now we can see that if we have a worldsheet wrapping near $ \rho = \infty$ at 
the angle $\psi_0$ then this worldsheet can be contracted to the origin with no change
in flux, since now the radial component of $H$ projected onto the worldsheet is proportional to
$\sin (\psi-\psi_0)$.
 But the flux of this worldsheet is $ \theta_{FT} - N \psi_0 $. It is this flux
that should be a multiple of $ 2 \pi$ so we see that we have $N$ different solutions
corresponding to $ \psi_0 = { \theta_{FT} \over N } + { 2 \pi n \over N}$ with
$ 0 \leq n < N$. 

Let us summarize this discussion. {}From the purely metric point of view, all the solutions
with arbitrary values of $\psi_0$ are non-singular, but once we consider the 
$B$ fields we see that only $N$ of the solutions are non-singular. 

It is easy to see what the gravity dual of a domain wall separating
two vacua is. Physicaly  we expect it to be something localized near $\rho \sim 0$ since
the theory is the same in the UV on both sides of the domain wall. But when we cross the 
domain wall we get two different solutions with different values of $\psi_0$ and 
we get a change in the flux of $B$ over the contractible sphere by $k$
units if 
$ \Delta \psi_0 = { 2 \pi k \over N}$. This implies that the domain wall should 
be $k$  NS 5 branes wrapping $S^3$.  

It is also possible to see how we can make $N$ of those fivebranes disappear. This is easier
to see from the seven dimensional point of view. We said that the seven dimensional theory
in the variables of \cite{Townsend:1983kk}
 has a three form potential. The fivebranes wrapped on $S^3 $
are electrically charged under this three form potential. In the seven dimensional lagrangian
there is a coupling of the form 
\eqn{coupl}{
i N \int A_3 \wedge Tr( F \wedge F)
}
where $F$ is the field strength  for the $SU(2)_R$ gauge fields. So we see that if 
we have $N$ fivebranes we can replace them by an instanton of the $SU(2)$ gauge field and
then expanding the instanton to infinite size we see that this kind of domain wall can
disappear.  This effect is of course familiar in the heterotic string context where we can
transform an NS fivebrane into an instanton in the gauge group
 \cite{Callan:1991dj}.
 In that case
one fivebrane was the same as one instanton.

\section{ Towards the pure ${\cal N} =1 $ theory}

If we intend to decouple the four dimensional theory we will have to take  
a limit where we go to scales much lower than the little string mass scale.
As shown in \cite{Itzhaki:1998dd} we need  to S-dualize the
gravity solution 
and switch to a D-fivebrane description.

The S-dual metric to that in \soltenns\ is 
\eqn{soltend}{\eqalign{
ds^2_{str} = & e^{\phi_D } \left[ dx^2_4 + N( d \rho^2 + e^{ 2 g(\rho)} d\Omega_2^2 + 
{1 \over 4 } \sum_a (w^a - A^a)^2 ) \right]
\cr
e^{2\phi_D} =& e^{2\phi_{D,0}}{\sinh 2 \rho \over
 2 e^{g(\rho)} } 
}}
and the NS $H$ field becomes a RR $H$ field. 
Everything that we said in the previous section about worldsheet instantons translates
into D-string instantons.

In this description an 
 external quark is a fundamental string that comes in  from infinity. When we have
a quark anti-quark pair and we separate them by a large distance  we see that
we find 
 a finite string tension from the point of view of the four dimensional theory
equal to 
\eqn{stringten}{
T_s = { e^{\phi_{D,0}} \over 2 \pi \alpha'}
}
The masses of glueballs  and Kaluza-Klein states on the spheres is, in the supergravity
approximation,  
\eqn{masses}{
M^2_{glueballs} \sim M^2_{KK} \sim { 1 \over N \alpha' }
}
Finally the tension of a domain wall interpolating between the $n$th and $n+1$th vacua, which
is now a D5 brane,  is
\eqn{domainwall}{
T_{wall} \sim { N^{3/2} e^{ 2 \phi_{D,0}}  }
}
Fundamental strings can end on these domain walls \cite{Witten:1998jd}. 
The baryon vertex is a D3 brane wrapped on $S^3$. 
A magnetic monopole source is a D-3 brane wrapping 
the sphere that 
 the worldsheet 
instantons were wrapping in the previous section and extending in the radial
and time  directions. 
They are screened because, since the sphere is contractible, each member of a monopole
anti-monopole pair can be wrapped in the three dimensional space parametrized by 
$\rho$ and the contractible sphere.

We see that in order to decouple the scale of the string tension from the 
scale of the KK states we need $ e^{\phi_{D,0}} N  \ll 1$.
This goes beyond the gravity
approximation, which requires $e^{\phi_{D,0}} N \gg 1$, 
 but it seems that we could still use this metric to formulate 
a string theory. This string theory should be such that it essentially has
no
excitations on $S^2 $ or $ S^3$. This is plausible since the sizes of those spheres is
smaller than the string scale. Presumably we should be able to replace the
six dimensional part of the geometry by a Liouville-like theory; since
this
geometry is  similar to the near conifold geometry, this sounds plausible.
In fact, for  
the near conifold geometry it was suggested 
that the sigma model can be replaced, for some calculations, 
by the $c=1$ (super) Liouville theory.
Indeed, in \cite{Ghoshal:1995wm} it was proved that the CFT is a Kazama-
Suzuki $SL(2)/U(1)$ coset model with level $k=3$. This theory was studied
in the context of non-critical bosonic strings in ref.\cite{Mukhi:1993zb}
and the  relation between  bosonic strings and SCFT of a conifold agrees 
with the results of
\cite{Witten:1992zd}.

It would be nice to understand the mapping to 
a Liouville-like theory in the case that we have RR fields. 
A nice feature of this RR sigma model is that it seems possible to choose 
light cone gauge. In AdS it is hard to choose light cone gauge, because
 in Poincare coordinates
we have a horizon. In this case there is no horizon and the light cone
 theory should be
better defined.  In the purely four dimensional theory we do not 
expect to have any 
dimensionless parameter. In our case we have a dimensionless parameter which is $\phi_0$, 
this parameter is related to the ratio of the QCD string tension (or mass scale) and
the six dimensional gauge coupling, or six dimensional scale of the little string theory. 
Presumably once we exchange the spheres by a Liouville theory we would find that
the string coupling is fixed in the IR  and of order ${1/N}$. 

Another related point is the precise coefficient for the beta function. 
In the 5-brane theory it is natural to define the scale as $g_{00}$ in D-string metric, 
since that will be the energy of a massive string mode sitting at position $\rho$. 
This gives a relation between the scale in the field theory and the position $\rho$ 
of the form $ \mu \sim e^{ \rho/2} $. When we look at the definition of the 
four dimensional string coupling  in \coup\ we see that 
$ 1/(g_4^2 N) \sim \log \mu/\Lambda_{QCD} $. But the coefficient is not the correct one. 
It is interesting that if we go to the five dimensional Einstein frame metric and we
define the scale as $ \mu^2 \sim g_{00}^{5,E} $ then we get precisely the right 
$\beta $ function with the right numerical  coefficient 
\cite{Intriligator:1996au}.
 We could not find any precise reason for 
choosing this UV/IR relation. In order to determine the precise relation it seems that
we should know the precise string theory and sigma model. 

In summary, this solution seems to  provide  a starting point for constructing the large N limit
of pure ${\cal N} = 1 $ Yang-Mills. We expect that the $S^3$ and $S^2$ would disappear
from the sigma model, leaving only the radial direction, and probably  also an angular 
direction, representing the $U(1)$ symmetry.  The final picture would have the
flavor of that in 
\cite{Aharony:1998ub},
 but it seems crucial to have  RR fields in order
to generate a warp factor in string frame.

\section*{Acknowledgements}

We would like to thank O. Aharony,  I. Klebanov, J. Polchinski,  M. Strassler,
A. Strominger  and C. Vafa
for discussions. 

The research of C.N. was supported by CONICET.
The research of J.M.\
was supported in part by DOE grant DE-FGO2-91ER40654,
NSF grant PHY-9513835, the Sloan Foundation and the
David and Lucile Packard Foundation.
We also thank the Aspen Center for Physics where part of this work was done.


\bibliography{none}
\bibliographystyle{ssg}

\end{document}